# Vector spherical quasi-Gaussian vortex beams

F.G. Mitri*

*Chevron – Area 52 Technology, Santa Fe, NM 87508, USA*

Model equations for describing and efficiently computing the radiation profiles of tightly spherically-focused higher-order electromagnetic beams of vortex nature are derived stemming from a vectorial analysis with the complex-source-point method. This solution, termed as a high-order quasi-Gaussian (qG) vortex beam, exactly satisfies the vector Helmholtz and Maxwell's equations. It is characterized by a nonzero integer degree and order ($n,m$), respectively, an arbitrary waist $w_0$, a diffraction convergence length known as the Rayleigh range $z_R$, and an azimuthal phase dependency in the form of a complex exponential corresponding to a vortex beam. An attractive feature of the high-order solution is the rigorous description of strongly focused (or strongly divergent) vortex wave-fields without the need of neither the higher-order corrections nor the numerically intensive methods. Closed-form expressions and computational results illustrate the analysis and some properties of the high-order qG vortex beams based on the axial and transverse polarization schemes of the vector potentials with emphasis on the beam waist.

PACS number(s): 41.20.Jb, 42.25.Bs, 42.15.Dp, 42.25.Fx, 02.90.+p

## I. INTRODUCTION

Modeling the beam-forming [1] of tightly focused beams [2-6] is a subject of particular interest in electromagnetic (EM)/optical/acoustical research, which received significant attention in the development of imaging microscopes, and other devices for particle manipulation and medical imaging. Usually, the predictions using numerical integration methods (commonly performed by means of the angular spectrum method of plane waves [7-10]) are computationally intensive, requiring the evaluation of a double integration procedure that can be time-consuming. Higher-order corrections [11-13] have been also suggested, which may provide an approximate solution with minimal numerical errors if a set of parameters is carefully chosen. Nevertheless, the lack of an exact solution for the description of tightly focused beams without any approximation, provided the impetus to further extend a method based on the complex source point (CSP) formalism [14-20] (Note the misprint in Eq.(3) of [20]; as written, Eq.(3) is not a proper solution of the Helmholtz equation since the angle $\theta$ is real as given in Eq.(2). The polar angle in Eq.(3) should have been expressed as a complex angle, $\theta_-$ given after Eq.(1) in the following), and introduce a solution corresponding to a fundamental (lowest-order) quasi-Gaussian (qG) beam, that is an exact solution of the Helmholtz equation and Maxwell's equations [21].

In this work, a *generalized* spherical vectorial solution of *vortex* nature that encompasses the lowest-order result [21] (see also the cylindrical counterpart solution in [22]), is provided, for which the degree and order ($n,m$) = (0,0). Vector solutions, which take into account the vector character of the field and its polarization, are necessary for the description of EM fields [23], especially when the wavelength is in the order of the beam waist. Moreover, it is of particular importance to develop *exact* vortex solutions that are applicable to the computation of tightly focused (or strongly divergent) wave-fields without any approximations nor the need for the higher-order corrections usually required for Gaussian beams [11, 12, 24], particularly for the computation of the beam-shape coefficients (BSCs) [25] used to obtain *a priori* information on the arbitrary scattering [26-29], radiation force [30, 31], orbital and angular spin momentum [32], and torque [30] in particle sizing, particle manipulation and optical tweezers to name a few applications. Unlike the lowest-order result [21], the present solutions are fundamentally different as they carry an angular momentum [32, 33] which sets a particle or a collection of particles into rotation by inducing a radiation torque [30]. Although previous vectorial analyses of vortex [28, 34-37] (and non-vortex [38-40]) beams of Bessel type have been developed, the beams were considered ideally non-diffracting in an unbounded space, or in other words of infinite extent.

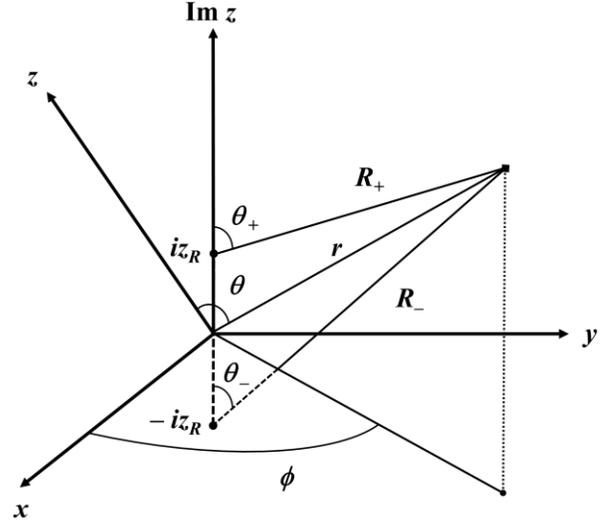

Fig. 1. Geometry of the problem.

---

*Emails: mitri@chevron.com, fmitri@gmail.com

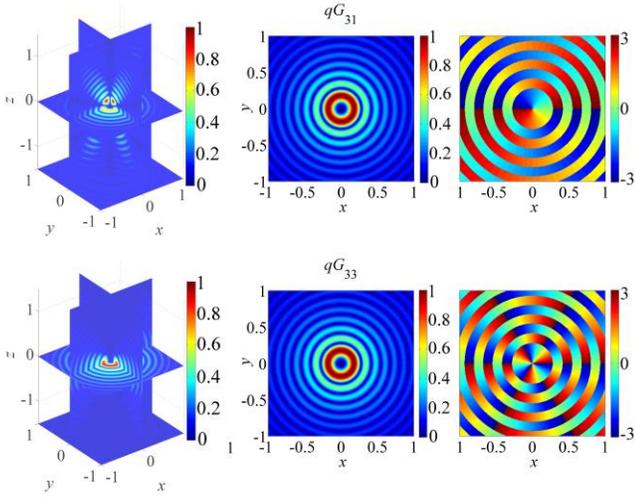

Fig. 2. (Color online) Axial (along the direction z) and cross-sectional (x,y- plane) magnitude and phase plots for qG$_{31}$ (1$^{st}$ row) and qG$_{33}$ (2$^{nd}$ row) for $kw_0 = 0.1$, corresponding to a strongly focused (or strongly divergent) beam. The units along the axes are in mm.

The present analysis starts by considering the general solution to the Helmholtz equation in spherical coordinates based on the method of separation of variables (i.e. Eq.(42) in [18]). The same method can be also applied using the CSP formalism (Eq.(22) in [16]) [14, 15, 18, 19, 41-43], and the result remains an exact solution of the Helmholtz equation. The effect of having the description of a generalized solution in a complex coordinates system, which may appear at first glance a simple artifice, has a major physical meaning in the description of evanescent waves [44] and the production of *finite* directional beams [17]. It is noted that the solutions presented in [16, 18] can be interpreted as a generalized set of spherical harmonics centered on a CSP. However, such solutions are singular at the CSP and may not be used to describe a physically realizable wave-field.

The removal of this singularity can be accomplished by introducing a sink in addition to the point source [45, 46], the sink being identical in amplitude and opposite in sign [47]. The vectorial solution, termed here a spherical "quasi-Gaussian" (qG) *vortex* beam, to make a distinction from the paraxial Gaussian solution used in conventional laser beams, consists of products of (nonsingular) spherical Bessel functions of the first kind and associated Legendre functions with a complex exponential phase dependency on the azimuthal angle.

## II. METHOD

A magnetic vector potential field $\mathbf{A}_{p,x}$ describing an exact solution of the vector wave equation (i.e. the source-free Helmholtz equation), and polarized along the *x*-direction is defined such that [47],

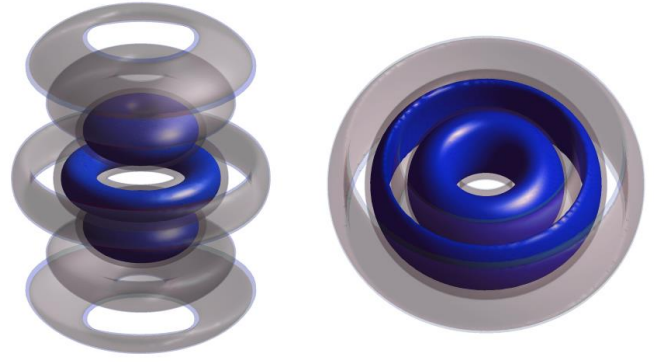

Fig. 3. (Color online) Isosurface plots corresponding to a qG$_{31}$ vortex beam of 1$^{st}$ order (left panel) and a qG$_{33}$ vortex beam of 3$^{rd}$ order (right panel) with a "doughnut" shape.

$$\mathbf{A}_{p,x} = qG_{nm}^{vortex}\mathbf{x},$$
$$= A_0 e^{\pm kz_R} j_n(\kappa_\pm) P_n^{|m|}(\cos\theta_\pm) e^{i|m|\phi}\mathbf{x}, \quad (1)$$

where the time-dependence in the form of $e^{-i\omega t}$ is suppressed from Eq.(1) for convenience, $\mathbf{x}$ is the unitary vector along the transverse *x*-direction, $A_0$ is the characteristic amplitude, $e^{\pm kz_R}$ is a normalization constant, the parameter $\kappa_\pm = kR_\pm$, $j_n(\cdot)$ is the spherical Bessel function of the first kind, $P_n^{|m|}(\cdot)$ are the associated Legendre functions of integer degree *n* and order *m*, $R_\pm = \sqrt{x^2 + y^2 + Z_\pm^2}$, $\theta_\pm = \cos^{-1}(Z_\pm/R_\pm)$, $\phi = \tan^{-1}(y/x)$, $Z_\pm = (z \pm z_0)$, and $z_R = kw_0^2/2$ where $w_0$ is the beam waist, $z_0 = iz_R$, and $z_R$ is the Rayleigh range (Fig. 1).

The analysis with complex angles introduces a representation for the field's characteristics, which allows determining the direction of propagation as well as field attenuation interpretations [44]. Moreover, the complex distance function $R_\pm$ is multi-valued (in this case four-valued) with branch point singularities on the circle defined by $\{x^2 + y^2 = z_R^2, z = 0\}$. A branch line (or cut) has to be introduced to make it single-valued [48]. Here, the branch cut is chosen such that $\{x^2 + y^2 \leq z_R^2, z = 0\}$, for which the complex distance function is continuous at all points except the branch cut.

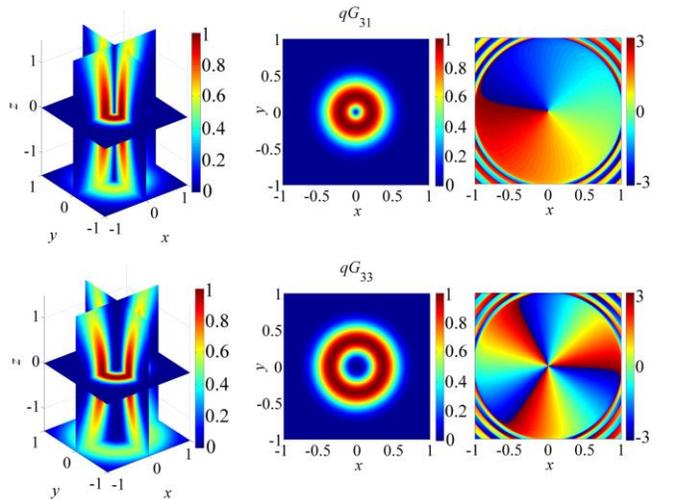

Fig. 4. (Color online) The same as in Fig. 2, however $kw_0 = 7$.

It is noted that the azimuthal dependence in Eq.(1) is expressed under the form of a complex exponential function which represents a vector potential of vortex nature. Eq.(1) is an exact solution of the vector Helmholtz equation, and the introduction of a sink along with an appropriate choice of the branch cut such that $\{x^2 + y^2 \leq z_R^2, z = 0\}$, makes this particular expression free from any singularities at $R_\pm = 0$. Unlike the spherical Hankel functions, the spherical Bessel functions are finite at $R_\pm = 0$. The cost of this choice, however, is that the CSP beam as given by Eq.(1), which propagates along the +/− z-direction, respectively, possesses a weak field component propagating backwards in the −/+ z-direction, respectively [45].

To illustrate this type of vortex beams, the magnitude and phase profiles of the vector potential given by Eq.(1) are computed for two given pairs $(n,m) = (3,1)$ and $(3,3)$, respectively. Two examples are chosen for which the value of $kw_0$ are selected to be $kw_0 = 0.1$, corresponding to a tightly focused (or strongly divergent) beam, and $kw_0 = 7$, corresponding to a quasi-collimated beam. The parameter $k = 25 \times 10^3 \text{m}^{-1}$, and the axial $z$ and transverse $(x,y)$ coordinates are varied by increments of $\delta(x,y,z) = 10^{-3}$ mm.

The panels in Fig. 2 show the comparison of both magnitude and phase profiles for qG beams of 3rd degree ($n = 3$) and 1st ($m = 1$) and 3rd ($m = 3$) orders, for a tightly focused (or tightly divergent) beam [i.e. $kw_0 = 0.1$]. The vortex nature of the beam is clearly manifested in the phase plot (3rd column) that varies in the cross-sectional plane according to the order of the beam. Moreover, the magnitude plots displayed in the central panels (i.e. cross-sectional plane $(x,y)$) show close similarity. On the other hand, the axial plots (1st column) show quite distinct features of the qG vortex beams. To better visualize the features, isosurface plots corresponding to the qG vortex beam of 3rd degree and 1st order (qG$_{31}$; 1st column, 1st row in Fig. 2) and the 3rd degree and 3rd order (qG$_{33}$; 1st column, 2nd row in Fig. 2) are displayed in Fig. 3.

The effect of changing the size parameter $kw_0 = 7$, which corresponds to quasi-collimated beams, is displayed in Fig. 4. One clearly notices the difference in the beam shape by comparing the results with Figs. 2 and 3, as well as the diameter increase of the hollow region when the order of the beam increases. This behavior has been previously observed for high-order Bessel vortex beams [49].

It is however important to note that the EM field has an *intrinsic vector* structure. Thus, for a complete description of the qG vortex beams, an electric and magnetic field should be defined using the vector potential given by Eq.(1) to account for the vector nature of the waves.

Using Lorenz' gauge condition [50], a magnetic field $\mathbf{H}_p$ is defined as,

$$\nabla \times \mathbf{A}_{p,x} = \mathbf{H}_p \varepsilon^{-1/2}. \qquad (2)$$

where $\varepsilon$ is the dielectric constant of the medium.

Thus, from Maxwell's equations and Eq.(2), the electric field is expressed as,

$$\mathbf{E}_p = ik\left[\mathbf{A}_{p,x} + \nabla\left(\nabla \cdot \mathbf{A}_{p,x}\right)/k^2\right]. \qquad (3)$$

Substituting Eq.(1) into Eqs.(2) and (3) leads to the determination of the electromagnetic field components. However, one notices an asymmetry in the mathematical expressions of the electric and magnetic field components, inappropriate in the physical description of symmetric beams. Therefore, the general physical description of beams in free space (with no imposed boundary conditions) requires using the dual field setup [42] to produce symmetrical behaviors in the EM field's components. The mathematical operation using the dual field consists of defining another electric vector potential $\mathbf{A}_{q,y}$ polarized along the negative (or positive) transverse y-direction such that,

$$\mathbf{A}_{q,y} = -qG_{nm}^{vortex}\mathbf{y}, \qquad (4)$$

where $\mathbf{y}$ is the unitary vector along the y-direction (Note here that the dual field setup procedure may not be required for the description of transverse modes [51] with specific boundary conditions imposed on the wave-field).

An electric field $\mathbf{E}_q$ may be therefore defined as,

$$\nabla \times \mathbf{A}_{q,y} = \mathbf{E}_q. \qquad (5)$$

Thus, from Maxwell's equations and Eq.(5), the magnetic field is expressed as,

$$\mathbf{H}_q \varepsilon^{-1/2} = -ik\left[\mathbf{A}_{q,y} + \nabla\left(\nabla \cdot \mathbf{A}_{q,y}\right)/k^2\right]. \qquad (6)$$

In the final mathematical procedure, the solution of Eqs.(2) and (3) is added to the solution of Eqs.(5) and (6), and dividing the end result by two, leads to the spatial Cartesian components for a quasi-Gaussian vortex beam of any integer degree $n$ and order $m$, which are expressed as,

$$E_{nm,x}^{(x,-y)} = \frac{A_0}{4} e^{(i|m|\phi \pm kz_R)} \left\{ \frac{2 j_n(\kappa_\pm) F_{nm}}{R_\pm^2} + \frac{Z_\pm P_n^{|m|}(\cos\theta_\pm) \Psi_n}{R_\pm^2} + 2ik \left[ j_n(\kappa_\pm) P_n^{|m|}(\cos\theta_\pm) + \frac{1}{k^2} \begin{pmatrix} \dfrac{j_n(\kappa_\pm) Z_\pm^2 x^2 [2Z_\pm F_{nm} + \Lambda_{nm}]}{R_\pm^4 (Z_\pm^2 - R_\pm^2)^2} + \dfrac{j_n(\kappa_\pm) Z_\pm F_{nm}}{R_\pm^2 (Z_\pm^2 - R_\pm^2)} \\ + \dfrac{2ixy|m| j_n(\kappa_\pm) Z_\pm F_{nm}}{R_\pm^2 (R_\pm^2 - Z_\pm^2)(x^2 + y^2)} + \dfrac{x^2 Z_\pm F_{nm}[3 j_n(\kappa_\pm) - \Psi_n]}{R_\pm^4 (R_\pm^2 - Z_\pm^2)} \\ + \dfrac{|m| y [2ix - |m| y] j_n(\kappa_\pm) P_n^{|m|}(\cos\theta_\pm)}{(x^2+y^2)^2} + \dfrac{P_n^{|m|}(\cos\theta_\pm) \Psi_n}{2R_\pm^2} - \dfrac{i|m| xy P_n^{|m|}(\cos\theta_\pm) \Psi_n}{R_\pm^2 (x^2+y^2)} \\ + \dfrac{x^2 P_n^{|m|}(\cos\theta_\pm)}{2R_\pm^4} \left[\dfrac{\xi_n}{2} - \Psi_n\right] \end{pmatrix} \right] \right\},$$

(7)

$$E_{nm,y}^{(x,-y)} = \frac{iA_0}{8k} e^{(i|m|\phi \pm kz_R)} \left\{ \frac{4 j_n(\kappa_\pm) Z_\pm^2 xy [2Z_\pm F_{nm} + \Lambda_{nm}]}{R_\pm^4 (Z_\pm^2 - R_\pm^2)^2} + \frac{4i|m| j_n(\kappa_\pm) Z_\pm F_{nm} [y^2 - x^2]}{R_\pm^2 (R_\pm^2 - Z_\pm^2)(x^2+y^2)} + \frac{4xy Z_\pm F_{nm}[\Psi_n - 3 j_n(\kappa_\pm)]}{R_\pm^4 (Z_\pm^2 - R_\pm^2)} - \frac{4i|m| j_n(\kappa_\pm) P_n^{|m|}(\cos\theta_\pm)}{(x^2+y^2)} \right. $$
$$\left. + \frac{4|m| j_n(\kappa_\pm) P_n^{|m|}(\cos\theta_\pm) y[|m|xy + 2iy]}{(x^2+y^2)^2} + \frac{2i|m| P_n^{|m|}(\cos\theta_\pm) \Psi_n [x^2 - y^2]}{R_\pm^2 (x^2+y^2)} + \frac{xy P_n^{|m|}(\cos\theta_\pm) [\xi_n - 2\Psi_n]}{R_\pm^4} \right\},$$

(8)

$$E_{nm,z}^{(x,-y)} = \frac{A_0}{8} e^{(i|m|\phi \pm kz_R)} \left\{ \frac{4iy|m| j_n(\kappa_\pm) P_n^{|m|}(\cos\theta_\pm)}{(x^2+y^2)} - \frac{4x Z_\pm j_n(\kappa_\pm) F_{nm}}{R_\pm^2 (Z_\pm^2 - R_\pm^2)} - \frac{2x P_n^{|m|}(\cos\theta_\pm) \Psi_n}{R_\pm^2} + \frac{i}{k} \begin{bmatrix} \dfrac{4 j_n(\kappa_\pm) Z_\pm x [Z_\pm F_{nm} - \Lambda_{nm}]}{R_\pm^4 (R_\pm^2 - Z_\pm^2)} + \dfrac{2Z_\pm^2 x F_{nm} \Psi_n}{R_\pm^4 (Z_\pm^2 - R_\pm^2)} \\ - \dfrac{2i|m| y P_n^{|m|}(\cos\theta_\pm) Z_\pm \Psi_n}{R_\pm^2 (x^2+y^2)} + \dfrac{x P_n^{|m|}(\cos\theta_\pm) Z_\pm [\xi_n - 2\Psi_n]}{R_\pm^4} \\ + \dfrac{4 j_n(\kappa_\pm) x F_{nm}}{R_\pm^2 (Z_\pm^2 - R_\pm^2)} - \dfrac{4iy|m| j_n(\kappa_\pm) F_{nm}}{R_\pm^2 (x^2+y^2)} + \dfrac{2x F_{nm} \Psi_n}{R_\pm^4} \end{bmatrix} \right\}$$

(9)

$$H_{nm,x}^{(x,-y)} = E_{nm,y}^{(x,-y)} \sqrt{\varepsilon},$$

(10)

$$H_{nm,y}^{(x,-y)} = \frac{A_0 \sqrt{\varepsilon}}{4} e^{(i|m|\phi \pm kz_R)} \left\{ \frac{2 j_n(\kappa_\pm) F_{nm}}{R_\pm^2} + \frac{Z_\pm P_n^{|m|}(\cos\theta_\pm) \Psi_n}{R_\pm^2} + 2ik \left[ j_n(\kappa_\pm) P_n^{|m|}(\cos\theta_\pm) + \frac{1}{k^2} \begin{pmatrix} \dfrac{j_n(\kappa_\pm) Z_\pm^2 y^2 [2Z_\pm F_{nm} + \Lambda_{nm}]}{R_\pm^4 (Z_\pm^2 - R_\pm^2)^2} + \dfrac{j_n(\kappa_\pm) Z_\pm F_{nm}}{R_\pm^2 (Z_\pm^2 - R_\pm^2)} - \dfrac{2ixy|m| j_n(\kappa_\pm) Z_\pm F_{nm}}{R_\pm^2 (R_\pm^2 - Z_\pm^2)(x^2+y^2)} \\ + \dfrac{y^2 Z_\pm F_{nm}[3 j_n(\kappa_\pm) - \Psi_n]}{R_\pm^4 (R_\pm^2 - Z_\pm^2)} - \dfrac{|m| x [2iy + |m| x] j_n(\kappa_\pm) P_n^{|m|}(\cos\theta_\pm)}{(x^2+y^2)^2} \\ + \dfrac{P_n^{|m|}(\cos\theta_\pm) \Psi_n}{2R_\pm^2} + \dfrac{i|m| xy P_n^{|m|}(\cos\theta_\pm) \Psi_n}{R_\pm^2 (x^2+y^2)} \\ + \dfrac{y^2 P_n^{|m|}(\cos\theta_\pm)}{2R_\pm^4} \left[\dfrac{\xi_n}{2} - \Psi_n\right] \end{pmatrix} \right] \right\},$$

(11)

$$H_{nm,z}^{(x,-y)} = \frac{A_0 \sqrt{\varepsilon}}{8} e^{(i|m|\phi \pm kz_R)} \left\{ -\frac{4ix|m| j_n(\kappa_\pm) P_n^{|m|}(\cos\theta_\pm)}{(x^2+y^2)} - \frac{4y Z_\pm j_n(\kappa_\pm) F_{nm}}{R_\pm^2 (Z_\pm^2 - R_\pm^2)} - \frac{2y P_n^{|m|}(\cos\theta_\pm) \Psi_n}{R_\pm^2} + \frac{i}{k} \begin{bmatrix} \dfrac{4 j_n(\kappa_\pm) Z_\pm y [Z_\pm F_{nm} - \Lambda_{nm}]}{R_\pm^4 (R_\pm^2 - Z_\pm^2)} + \dfrac{2Z_\pm^2 y F_{nm} \Psi_n}{R_\pm^4 (Z_\pm^2 - R_\pm^2)} \\ + \dfrac{2i|m| x P_n^{|m|}(\cos\theta_\pm) Z_\pm \Psi_n}{R_\pm^2 (x^2+y^2)} + \dfrac{y P_n^{|m|}(\cos\theta_\pm) Z_\pm [\xi_n - 2\Psi_n]}{R_\pm^4} \\ + \dfrac{4 j_n(\kappa_\pm) y F_{nm}}{R_\pm^2 (Z_\pm^2 - R_\pm^2)} + \dfrac{4ix|m| j_n(\kappa_\pm) F_{nm}}{R_\pm^2 (x^2+y^2)} + \dfrac{2y F_{nm} \Psi_n}{R_\pm^4} \end{bmatrix} \right\},$$

(12)

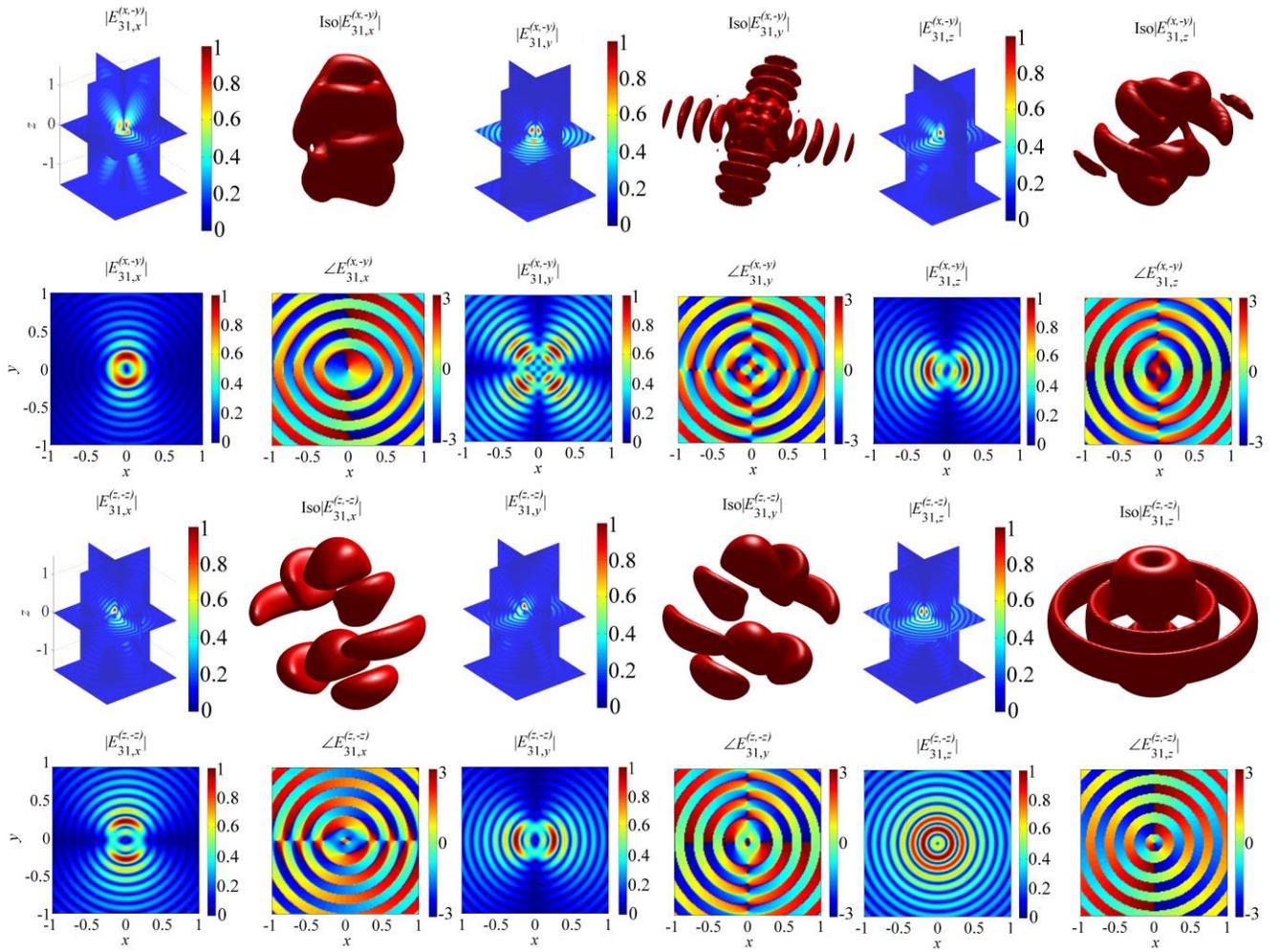

Fig. 5. (Color online) Comparison between the axial magnitude (along the direction $z$), isosurface and cross-sectional ($x,y$- plane) magnitude and phase plots for the electric components of a qG$_{31}$ vortex beam for $kw_0 = 0.1$, corresponding to a strongly focused (or strongly divergent) beam. The 1$^{st}$ and 2$^{nd}$ rows correspond to the ($x,-y$) configuration, whereas the 3$^{rd}$ and 4$^{th}$ rows correspond to the ($z,-z$) configuration. The units along the axes are in mm.

where the superscript $(x,-y)$ in the EM field's components [i.e. Eqs.(7)-(12)] denotes the polarization state of the vector potentials $\mathbf{A}_{p,x}$ and $\mathbf{A}_{q,y}$, respectively. Moreover, the functions appearing in these expressions are given by,

$$F_{nm} = \left[(n+1)Z_{\pm}P_n^{|m|}(\cos\theta_{\pm}) + (|m|-n-1)R_{\pm}P_{n+1}^{|m|}(\cos\theta_{\pm})\right],$$

$$\Psi_n = \left\{\kappa_{\pm}\left[j_{n-1}(\kappa_{\pm}) - j_{n+1}(\kappa_{\pm})\right] - j_n(\kappa_{\pm})\right\},$$

$$\Lambda_{nm} = \begin{cases} n(n+1)P_n^{|m|}(\cos\theta_{\pm})Z_{\pm}^2 + (|m|-n-1)(2n+3)R_{\pm}Z_{\pm}P_{n+1}^{|m|}(\cos\theta_{\pm}) \\ + R_{\pm}^2\left[(n+1)P_n^{|m|}(\cos\theta_{\pm}) + (|m|^2 - (2n+3)|m| + n^2 + 3n + 2)P_{n+2}^{|m|}(\cos\theta_{\pm})\right] \end{cases},$$

$$\xi_n = \left\{\kappa_{\pm}^2\left[j_{n-2}(\kappa_{\pm}) - 2j_n(\kappa_{\pm}) + j_{n+2}(\kappa_{\pm})\right] + 2\kappa_{\pm}\left[j_{n+1}(\kappa_{\pm}) - j_{n-1}(\kappa_{\pm})\right] + 3j_n(\kappa_{\pm})\right\}.$$

(13)

It has been also recognized that other states of polarization [52], such as the *axial* polarization scheme, can be particularly useful in the development of free-electron lasers [53-56]. Thus, the aim here is to further extend the analysis to the case where the vector potentials are polarized along the axial directions $\pm z$ by deriving closed-form expressions for the EM field's components in this configuration. Thus, a vector potential field $\mathbf{A}_{p,z}$ denoting an exact solution of the Helmholtz equation, and polarized along the $z$ direction is expressed as,

$$\mathbf{A}_{p,z} = q\mathrm{G}_{nm}^{vortex}\mathbf{z}, \quad (14)$$

where $q\mathrm{G}_{nm}^{vortex}$ is given by Eq.(1), and $\mathbf{z}$ is the unitary vector along the $z$-direction. Following the procedure as given by Eqs.(2)-(6) after defining another vector potential polarized along the negative axial direction as,

$$\mathbf{A}_{q,z} = -q\mathrm{G}_{nm}^{vortex}\mathbf{z}, \quad (15)$$

and manipulating the results, the spatial Cartesian components for a spherical quasi-Gaussian vortex beam of any integer degree $n$ and order $m$ in the $(z,-z)$ polarization scheme, are found to be,

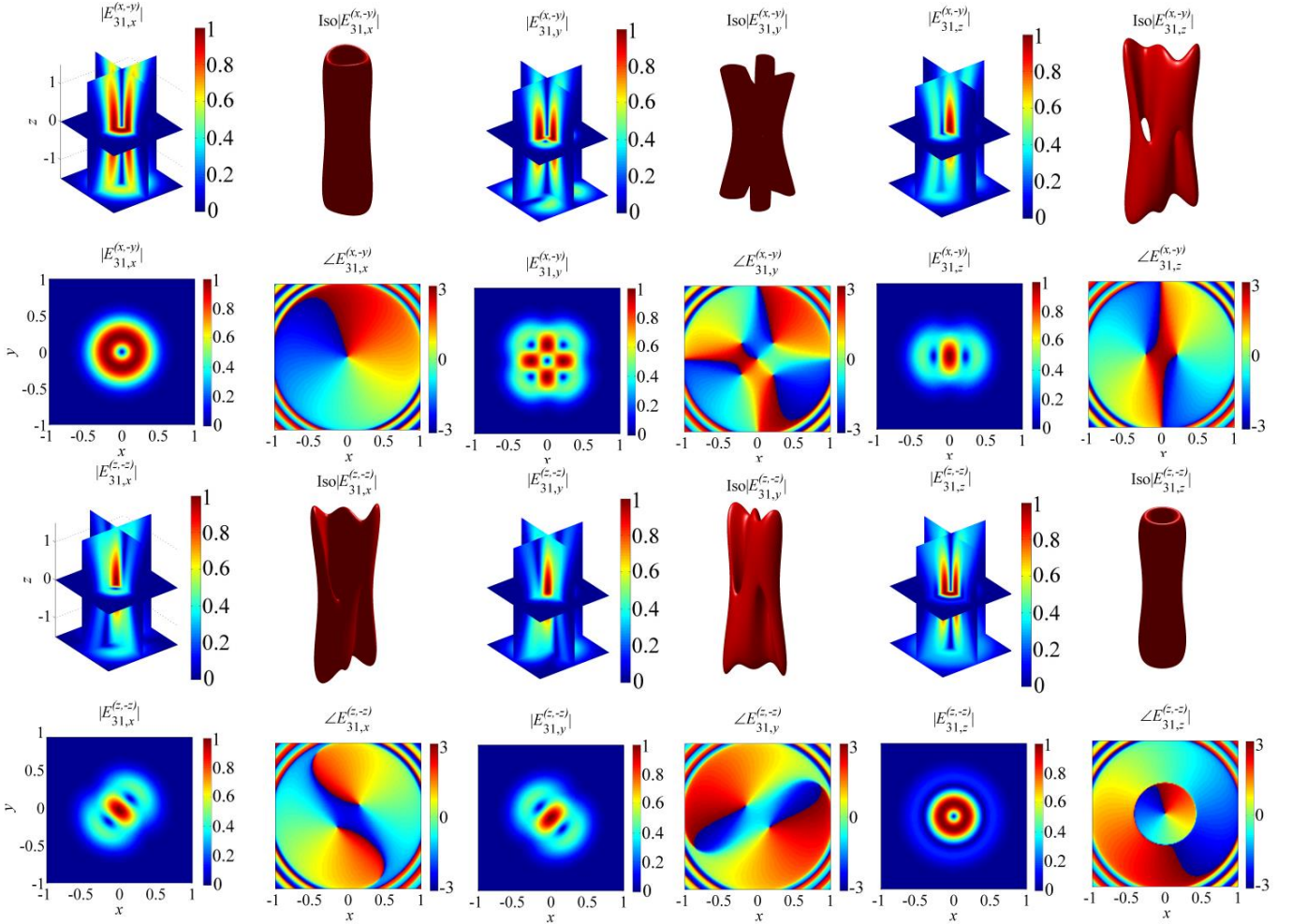

Fig. 6. (Color online) The same as in Fig. 5, but the size parameter is $kw_0 = 7$, corresponding to a quasi-collimated beam.

$$E_{nm,x}^{(z,-z)} = \frac{A_0}{8} e^{(i|m|\phi \pm kz_R)} \left\{ -\frac{4ix|m| j_n(\kappa_\pm) P_n^{|m|}(\cos\theta_\pm)}{(x^2+y^2)} - \frac{4yZ_\pm j_n(\kappa_\pm) F_{nm}}{R_\pm^2 (Z_\pm^2 - R_\pm^2)} - \frac{2yP_n^{|m|}(\cos\theta_\pm)\Psi_n}{R_\pm^2} + \frac{i}{k} \begin{bmatrix} \frac{2Z_\pm x \left[ 2j_n(\kappa_\pm)(2Z_\pm F_{nm} + \Lambda_{nm}) + Z_\pm F_{nm} \Psi_n \right]}{R_\pm^4 (Z_\pm^2 - R_\pm^2)} \\ -\frac{2i|m|y \left[ P_n^{|m|}(\cos\theta_\pm) Z_\pm \Psi_n + 2j_n(\kappa_\pm) F_{nm} \right]}{R_\pm^2 (x^2+y^2)} + \frac{2xF_{nm}\Psi_n}{R_\pm^4} \\ + \frac{xP_n^{|m|}(\cos\theta_\pm) Z_\pm [\xi_n - 2\Psi_n]}{R_\pm^4} + \frac{4x(3Z_\pm^2 - R_\pm^2) j_n(\kappa_\pm) F_{nm}}{R_\pm^4 (R_\pm^2 - Z_\pm^2)} \end{bmatrix} \right\}, \quad (16)$$

$$E_{nm,y}^{(z,-z)} = \frac{A_0}{8} e^{(i|m|\phi \pm kz_R)} \left\{ -\frac{4iy|m| j_n(\kappa_\pm) P_n^{|m|}(\cos\theta_\pm)}{(x^2+y^2)} + \frac{4xZ_\pm j_n(\kappa_\pm) F_{nm}}{R_\pm^2 (Z_\pm^2 - R_\pm^2)} + \frac{2xP_n^{|m|}(\cos\theta_\pm)\Psi_n}{R_\pm^2} + \frac{i}{k} \begin{bmatrix} \frac{2Z_\pm y \left[ 2j_n(\kappa_\pm)(2Z_\pm F_{nm} + \Lambda_{nm}) + Z_\pm F_{nm} \Psi_n \right]}{R_\pm^4 (Z_\pm^2 - R_\pm^2)} \\ + \frac{2i|m|x \left[ P_n^{|m|}(\cos\theta_\pm) Z_\pm \Psi_n + 2j_n(\kappa_\pm) F_{nm} \right]}{R_\pm^2 (x^2+y^2)} + \frac{2yF_{nm}\Psi_n}{R_\pm^4} \\ + \frac{yP_n^{|m|}(\cos\theta_\pm) Z_\pm [\xi_n - 2\Psi_n]}{R_\pm^4} + \frac{4y(3Z_\pm^2 - R_\pm^2) j_n(\kappa_\pm) F_{nm}}{R_\pm^4 (R_\pm^2 - Z_\pm^2)} \end{bmatrix} \right\}, \quad (17)$$

$$E_{nm,z}^{(z,-z)} = \frac{-iA_0}{2\kappa_\pm R_\pm^4 (|m|+n+1)} e^{(i|m|\phi \pm kz_R)} \left\{ \begin{array}{l} j_n(\kappa_\pm) \left[ \kappa_\pm^2 R_\pm^2 \chi_{nm} + (4n^2-1) Z_\pm^2 \chi_{nm} + R_\pm^2 \Phi_{nm} \right] \\ -\kappa_\pm j_{n+1}(\kappa_\pm) \left[ R_\pm^2 \Gamma_{nm} + (2n-1) Z_\pm^2 \chi_{nm} \right] \end{array} \right\}, \quad (18)$$

$$H_{nm,x}^{(z,-z)} = \frac{A_0 \sqrt{\varepsilon}}{8} e^{(i|m|\phi \pm kz_R)} \left\{ \frac{4ix|m| j_n(\kappa_\pm) P_n^{|m|}(\cos\theta_\pm)}{(x^2+y^2)} + \frac{4yZ_\pm j_n(\kappa_\pm) F_{nm}}{R_\pm^2 (Z_\pm^2 - R_\pm^2)} + \frac{2yP_n^{|m|}(\cos\theta_\pm)\Psi_n}{R_\pm^2} - \frac{i}{k} \begin{bmatrix} \frac{2Z_\pm x \left[ 2j_n(\kappa_\pm)(2Z_\pm F_{nm} + \Lambda_{nm}) + Z_\pm F_{nm} \Psi_n \right]}{R_\pm^4 (R_\pm^2 - Z_\pm^2)} \\ + \frac{2i|m|y \left[ P_n^{|m|}(\cos\theta_\pm) Z_\pm \Psi_n + 2j_n(\kappa_\pm) F_{nm} \right]}{R_\pm^2 (x^2+y^2)} - \frac{2xF_{nm}\Psi_n}{R_\pm^4} \\ -\frac{xP_n^{|m|}(\cos\theta_\pm) Z_\pm [\xi_n - 2\Psi_n]}{R_\pm^4} - \frac{4x(3Z_\pm^2 - R_\pm^2) j_n(\kappa_\pm) F_{nm}}{R_\pm^4 (R_\pm^2 - Z_\pm^2)} \end{bmatrix} \right\}, \quad (19)$$

$$H_{nm,y}^{(z,-z)} = \frac{A_0 \sqrt{\varepsilon}}{8} e^{(i|m|\phi \pm kz_R)} \left\{ \frac{4iy|m| j_n(\kappa_\pm) P_n^{|m|}(\cos\theta_\pm)}{(x^2+y^2)} - \frac{4xZ_\pm j_n(\kappa_\pm) F_{nm}}{R_\pm^2 (Z_\pm^2 - R_\pm^2)} - \frac{2xP_n^{|m|}(\cos\theta_\pm)\Psi_n}{R_\pm^2} - \frac{i}{k} \begin{bmatrix} \frac{2Z_\pm y \left[ 2j_n(\kappa_\pm)(2Z_\pm F_{nm} + \Lambda_{nm}) + Z_\pm F_{nm} \Psi_n \right]}{R_\pm^4 (R_\pm^2 - Z_\pm^2)} \\ -\frac{2i|m|x \left[ P_n^{|m|}(\cos\theta_\pm) Z_\pm \Psi_n + 2j_n(\kappa_\pm) F_{nm} \right]}{R_\pm^2 (x^2+y^2)} - \frac{2yF_{nm}\Psi_n}{R_\pm^4} \\ -\frac{yP_n^{|m|}(\cos\theta_\pm) Z_\pm [\xi_n - 2\Psi_n]}{R_\pm^4} - \frac{4y(3Z_\pm^2 - R_\pm^2) j_n(\kappa_\pm) F_{nm}}{R_\pm^4 (R_\pm^2 - Z_\pm^2)} \end{bmatrix} \right\}, \quad (20)$$

$$H_{nm,z}^{(z,-z)} = E_{nm,z}^{(z,-z)} \sqrt{\varepsilon}. \quad (21)$$

Moreover, the additional functions appearing in Eqs.(16)-(21) are expressed as,

$$\chi_{nm} = \left[ (|m|-n-1) Z_\pm P_{n+1}^{|m|}(\cos\theta_\pm) + \sqrt{R_\pm^2 - Z_\pm^2} P_{n+1}^{|m|+1}(\cos\theta_\pm) \right],$$

$$\Phi_{nm} = \begin{bmatrix} (|m|-n-1) Z_\pm (|m|^2 - 2|m|n + |m| - 3n^2 - k^2 Z_\pm^2 - n + 1) P_{n+1}^{|m|}(\cos\theta_\pm) \\ + (|m|^2 - n^2 - k^2 Z_\pm^2) \sqrt{R_\pm^2 - Z_\pm^2} P_{n+1}^{|m|+1}(\cos\theta_\pm) \end{bmatrix},$$

$$\Gamma_{nm} = \left[ (-2|m|^2 + |m| + 2n^2 + 3n + 1) Z_\pm P_{n+1}^{|m|}(\cos\theta_\pm) + \sqrt{R_\pm^2 - Z_\pm^2} P_{n+1}^{|m|+1}(\cos\theta_\pm) \right]. \quad (22)$$

To illustrate the vectorial analysis, the magnitude, isosurface profiles and phase plots of only the *electric* field's components in both the (x,–y) and (z,–z) configurations (i.e. Eqs. (7)-(9) and (16)-(18), respectively) are computed for the pair $(n,m) = (3,1)$. In the simulations, two values of the size parameter $kw_0$ are selected to be $kw_0 = 0.1$ (Fig. 5), corresponding to a tightly focused (or strongly divergent) beam, and $kw_0 = 7$ (Fig. 6), corresponding to a quasi-collimated beam. The parameter $k = 25 \times 10^3 \text{m}^{-1}$, and the axial $z$ and transverse $(x,y)$ coordinates are varied by increments of $\delta(x,y,z) = 10^{-3}$ mm.

Comparisons of the 1st with the 3rd row, and the 2nd with the 4th row in Fig. 5 clearly show the effect of changing the polarization states of the vector potentials from the transverse ($x,-y$) to the axial ($z,-z$) one. One particularly notices the spatial distributions in the cross-sectional plane for the components $\left|E_{31,x}^{(x,-y)}\right|$ and $\left|E_{31,z}^{(x,-y)}\right|$ (2nd row, 1st and 5th panel, respectively), which show an asymmetry in the central part of the plots, only in the transverse ($x,-y$) polarization state. The central area of the beam appears to be slightly rotated in the plane so that the symmetry is broken. This is a characteristic of the qG$_{31}$ vortex beam for $kw_0 = 0.1$.

This antisymmetry is removed when the beam becomes more directional as displayed in Fig. 6 (2nd row, 1st and 5th panel, respectively) for $kw_0 = 7$. Moreover, evolutions of the components $\left|E_{31,x}^{(z,-z)}\right|$ and $\left|E_{31,y}^{(z,-z)}\right|$ (4th row) are perceived; the null observed when $kw_0 = 0.1$ (Fig. 5, 4th row, 1st and 3rd panel) is transformed into a maximum in magnitude with a rotation in the transverse plane (Fig. 6, 4th row, 1st and 3rd panel).

As mentioned previously, the high-order qG vortex beams carry both linear and angular momenta, responsible for the production of a radiation force and torque [30] on a particle. A recent analysis dealing with a coherent superposition of Bessel beams [37] has shown that both linear and angular momentum density fluxes may reverse sign at particular values of the half-cone angle of the beam. These behaviors anticipate the production of a "tractor" beam where particulate matter may be pulled back toward the source, and a spinning reversal effect in which particulate matter may rotate with opposite handedness to the beam. For the present case of high-order qG vortex beams, the analysis [37] can be directly extended to evaluate the linear and angular momentum density fluxes with particular emphasis on the dimensionless waist $kw_0$, since the EM components of the qG vortex beam are now available (i.e. Eqs. (7)-(12), (16)-(21)). Further investigations focused on evaluating the EM radiation force and torque on a particle are beyond the scope of the present study, and will be the subject of future research.

### III. CONCLUSION

In summary, the vector wave properties for a CSP vortex solution representing tightly spherically-focused beams, are investigated. Particular emphasis is given on the polarizations of the electric and magnetic vector potentials, which produce distinct components for the EM field, with vortex behavior. In addition, the effect of increasing the beam waist produces quasi-collimated beams in the broad sense. The field's expressions are exact solutions of Maxwell's equations and are obtained without any approximations. Potential use of the solutions is in modeling strongly focused (or quasi-collimated) beams without the need of numerical integration procedures, nor the higher-order corrections. Other potential application is the accurate computation of the beam-shape coefficients used in the Generalized Lorenz-Mie Theories (GLMTs) for evaluating the arbitrary scattering, forces, and torques on particles using tightly focused laser vortex beams.

[1] A. Ciattoni, C. Conti, and P. Di Porto, Phys. Rev. E **69**, 036608 (2004).
[2] S. A. Self, Appl. Opt. **22**, 658 (1983).
[3] R. Dorn, S. Quabis, and G. Leuchs, Phys. Rev. Lett. **91**, 233901 (2003).
[4] J. Lermé, G. Bachelier, P. Billaud, C. Bonnet, M. Broyer, E. Cottancin, S. Marhaba, and M. Pellarin, J. Opt. Soc. Am. A **25**, 493 (2008).
[5] J. Lermé, C. Bonnet, M. Broyer, E. Cottancin, S. Marhaba, and M. Pellarin, Phys. Rev. B **77**, 245406 (2008).
[6] N. M. Mojarad, V. Sandoghdar, and M. Agio, J. Opt. Soc. Am. B **25**, 651 (2008).
[7] E. Wolf, Proc. R. Soc. London Ser. A **253**, 349 (1959).
[8] B. Richards and E. Wolf, Proc. R. Soc. London Ser. A **253**, 358 (1959).
[9] P. Varga and P. Török, Opt. Comm. **152**, 108 (1998).
[10] A. Rohrbach and E. H. K. Stelzer, J. Opt. Soc. Am. A **18**, 839 (2001).
[11] J. P. Barton and D. R. Alexander, J. Appl. Phys. **66**, 2800 (1989).
[12] P. B. Bareil and Y. Sheng, J. Opt. Soc. Am. A **30**, 1 (2013).
[13] J. A. Lock, J. Quant. Spectr. Rad. Tran. **126**, 16 (2013).
[14] Y. A. Kravtsov, Radiophys. Quant. Electron. **10**, 719 (1967).
[15] G. A. Deschamps, Electron. Lett. **7**, 684 (1971).
[16] M. Couture and P. A. Belanger, Phys. Rev. A **24**, 355 (1981).
[17] L. B. Felsen, Geophys. J. R. Astr. Soc.iety **79**, 77 (1984).
[18] B. T. Landesman and H. H. Barrett, J. Opt. Soc. Am. A **5**, 1610 (1988).
[19] C. J. R. Sheppard and S. Saghafi, Phys. Rev. A **57**, 2971 (1998).
[20] S. Orlov and U. Peschel, Phys. Rev. A **82**, 063820 (2010).
[21] F. G. Mitri, Phys. Rev. A **87**, 035804 (2013).
[22] F. G. Mitri, Opt. Lett. **38**, 4727 (2013).
[23] J. A. Stratton and L. J. Chu, Phys. Rev. **56**, 99 (1939).
[24] L. W. Davis, Phys. Rev. A **19**, 1177 (1979).
[25] G. Gouesbet and G. Grehan, *Generalized Lorenz-Mie Theories* (Springer, Berlin, 2011), 1st edn.
[26] J. P. Barton, D. R. Alexander, and S. A. Schaub, J. Appl. Phys. **64**, 1632 (1988).
[27] F. G. Mitri, Opt. Lett. **36**, 766 (2011).
[28] F. G. Mitri, IEEE Trans. Antenn. Prop. **59**, 4375 (2011).
[29] J. A. Lock, S. Y. Wrbanek, and K. E. Weiland, Appl. Opt. **45**, 3634 (2006).
[30] J. P. Barton, D. R. Alexander, and S. A. Schaub, J. Appl. Phys. **66**, 4594 (1989).
[31] J. A. Lock, Appl. Opt. **43**, 2532 (2004).
[32] L. Allen, S. M. Barnett, and M. J. Padgett, edited by IOP Publishing (CRC Press, Hoboken, 2003).
[33] A. M. Yao and M. J. Padgett, Adv. Opt. Photon. **3**, 161 (2011).
[34] F. G. Mitri, Opt. Lett. **36**, 606 (2011).
[35] F. G. Mitri, Opt. Lett. **38**, 615 (2013).
[36] F. G. Mitri, Optik - Int. J. Light Electron Opt. **124**, 1469 (2013).
[37] F. G. Mitri, Phys. Rev. A **88**, 035804 (2013).
[38] F. G. Mitri, Wave Motion **49**, 561 (2012).
[39] F. G. Mitri, Phys. Rev. A **85**, 025801 (2012).
[40] F. G. Mitri, Eur. Phys. J. D **67**, 1, 135 (2013).
[41] J. B. Keller and W. Streifer, J. Opt. Soc. Am. **61**, 40 (1971).
[42] A. L. Cullen and P. K. Yu, Proc. R. Soc. London Ser. A. **366**, 155 (1979).
[43] C. J. R. Sheppard, J. Opt. Soc. Am. A **18**, 1579 (2001).
[44] L. B. Felsen, J. Opt. Soc. Am. **66**, 751 (1976).
[45] E. Heyman, B. Z. Steinberg, and L. B. Felsen, J. Opt. Soc. Am. A **4**, 2081 (1987).
[46] M. V. Berry, J. Phys. A **27**, L391 (1994).
[47] Z. Ulanowski and I. K. Ludlow, Opt. Lett. **25**, 1792 (2000).
[48] E. Heyman and L. B. Felsen, J. Opt. Soc. Am. A **18**, 1588 (2001).
[49] F. G. Mitri, Ann. Phys. (NY) **323**, 2840 (2008).
[50] J. D. Jackson, *Classical electrodynamics* (Wiley, 1999, p. 240), Classical electrodynamics, 3rd edition.
[51] A. April, Opt. Lett. **33**, 1563 (2008).
[52] J. Lekner, J. Opt. A **5**, 6 (2003).
[53] L. W. Davis and G. Patsakos, Opt. Lett. **6**, 22 (1981).
[54] J. Verbeeck, H. Tian, and P. Schattschneider, Nature **467**, 301 (2010).
[55] B. J. McMorran, A. Agrawal, I. M. Anderson, A. A. Herzing, H. J. Lezec, J. J. McClelland, and J. Unguris, Science **331**, 192 (2011).
[56] L. Clark, A. Béché, G. Guzzinati, A. Lubk, M. Mazilu, R. Van Boxem, and J. Verbeeck, Phys. Rev. Lett. **111**, 064801 (2013).